\documentclass[twocolumn,showpacs,preprintnumbers,amsmath,amssymb,prl]{revtex4}

\usepackage{graphicx}
\usepackage{dcolumn}
\usepackage{bm}
\usepackage{tabularx}
\usepackage{amsmath}

\begin{document}

\title{Robust nodal superconductivity induced by isovalent doping in\\ Ba(Fe$_{1-x}$Ru$_x$)$_2$As$_2$ and BaFe$_2$(As$_{1-x}$P$_x$)$_2$}

\author{X. Qiu,$^1$ S. Y. Zhou,$^1$ H. Zhang,$^1$ B. Y. Pan,$^1$ X. C. Hong,$^1$ Y. F. Dai,$^1$\\ Man Jin Eom,$^2$ Jun Sung Kim,$^2$ Z. R. Ye,$^1$ Y. Zhang,$^1$ D. L. Feng,$^1$ S. Y. Li$^{1,*}$}

\affiliation{$^1$Department of Physics, State Key Laboratory of Surface Physics, and Laboratory of Advanced Materials, Fudan University, Shanghai 200433, China\\
$^2$Department of Physics, Pohang University of Science and
Technology, Pohang 790-784, Korea}

\date{\today}

\begin{abstract}
We present the ultra-low-temperature heat transport study of
iron-based superconductors Ba(Fe$_{1-x}$Ru$_x$)$_2$As$_2$ and
BaFe$_2$(As$_{1-x}$P$_x$)$_2$. For optimally doped
Ba(Fe$_{0.64}$Ru$_{0.36}$)$_2$As$_2$, a large residual linear term
$\kappa_0/T$ at zero field and a $\sqrt{H}$ dependence of
$\kappa_0(H)/T$ are observed, which provide strong evidences for
nodes in the superconducting gap. This result demonstrates that the
isovalent Ru doping can also induce nodal superconductivity, as P
does in BaFe$_2$(As$_{0.67}$P$_{0.33}$)$_2$. Furthermore, in
underdoped Ba(Fe$_{0.77}$Ru$_{0.23}$)$_2$As$_2$ and heavily
underdoped BaFe$_2$(As$_{0.82}$P$_{0.18}$)$_2$, $\kappa_0/T$
manifests similar nodal behavior, which shows the robustness of
nodal superconductivity in the underdoped regime and puts constraint
on theoretical models.
\end{abstract}


\maketitle

Since the discovery of high-$T_c$ superconductivity in iron
pnictides \cite{Kamihara,Paglione}, the electronic pairing mechanism
has been a central issue \cite{FaWang1}. One key to understand it is
to clarify the symmetry and structure of the superconducting gap
\cite{Hirschfeld}. However, even for the most studied
(Ba,Sr,Ca,Eu)Fe$_2$As$_2$ (122) system, the situation is still
fairly complex \cite{Hirschfeld}.

Near optimal doping, for both hole- and electron-doped 122
compounds, the angle-resolved photon emission spectroscopy (ARPES)
experiments clearly demonstrated multiple nodeless superconducting
gaps \cite{HDing,KTerashima}, which was further supported by bulk
measurements such as thermal conductivity
\cite{XGLuo,LDing,Tanatar}. On the overdoped side, nodal
superconductivity was found in the extremely hole-doped
KFe$_2$As$_2$ \cite{JKDong1,KHashimoto1}, while strongly anisotropic
gap \cite{Tanatar}, or isotropic gaps with significantly different
magnitudes \cite{JKDong2,YBang} were suggested in the heavily
electron-doped Ba(Fe$_{1-x}$Co$_x$)$_2$As$_2$. On the underdoped
side, recent heat transport measurements claimed possible nodes in
the superconducting gap of hole-doped Ba$_{1-x}$K$_x$Fe$_2$As$_2$
with $x <$ 0.16 \cite{JReid1}, in contrast to the nodeless gaps
found in electron-doped Ba(Fe$_{1-x}$Co$_x$)$_2$As$_2$
\cite{Tanatar}.

Intriguingly, nodal superconductivity was also found in optimally
doped BaFe$_2$(As$_{0.67}$P$_{0.33}$)$_2$ ($T_c =$ 30 K)
\cite{YNakai,KHashimoto2}, in which the superconductivity is induced
by isovalent P doping. The ARPES experiments have given conflicting
results on the position of the nodes \cite{TShimojima,YZhang}.
Moreover, previously LaFePO ($T_c \sim$ 6 K) displays clear nodal
behavior \cite{Fletcher,Hicks,MYamashida}, and recently there is
penetration depth evidence for nodes in the superconducting gap of
LiFeP ($T_c \sim$ 4.5 K) \cite{KHashimoto3}. The nodal
superconductivity in these P-doped compounds are very striking,
which raises the puzzling question why the P doping is so special in
iron-based superconductors. The theoretical explanations of this
puzzle are far from consensus
\cite{KKuroki,FaWang2,RThomale,KSuzuki}.

\begin{figure}
\includegraphics[clip,width=4cm]{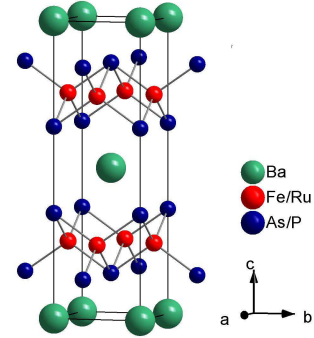}
\caption{(Color online). Isovalent doping in the Fe$_2$As$_2$ slabs
of BaFe$_2$As$_2$, by substituting As with P, or Fe with Ru. Both
ways can induce superconductivity, and result in similar phase
diagrams.}
\end{figure}

A recent proposal is that the nodal state in iron-pnictide
superconductors, except for KFe$_2$As$_2$, is induced when the
pnictogen height $h_{Pn}$ from the iron plane decreases below a
threshold value of $\sim$ 1.33 \AA\ \cite{KHashimoto3}. According to
this proposal, there may exist a transition from nodal to nodeless
state tuned by $h_{Pn}$, for example, in underdoped
BaFe$_2$(As$_{1-x}$P$_x$)$_2$. Therefore, it is important to
investigate the doping evolution of the superconducting gap
structure in BaFe$_2$(As$_{1-x}$P$_x$)$_2$. In another aspect, since
isovalent substituting Fe with Ru in BaFe$_2$As$_2$, as shown in
Fig. 1, can also decrease $h_{Pn}$ and result in similar phase
diagram \cite{SSharma,FRullier-Albenque,AThaler}, it will be very
interesting to check whether the gap of
Ba(Fe$_{1-x}$Ru$_x$)$_2$As$_2$ has nodes.

In this Letter, we report the demonstration of nodal
superconductivity in optimally doped
Ba(Fe$_{0.64}$Ru$_{0.36}$)$_2$As$_2$, underdoped
Ba(Fe$_{0.77}$Ru$_{0.23}$)$_2$As$_2$, and heavily underdoped
BaFe$_2$(As$_{0.82}$P$_{0.18}$)$_2$  by thermal conductivity
measurements down to 50 mK. Our finding of nodal gap in
Ba(Fe$_{0.64}$Ru$_{0.36}$)$_2$As$_2$ suggests a common origin of the
nodal superconductivity induced by isovalent P and Ru doping. The
nodal gap in Ba(Fe$_{0.77}$Ru$_{0.23}$)$_2$As$_2$ and
BaFe$_2$(As$_{0.82}$P$_{0.18}$)$_2$ shows no transition from nodal
to nodeless state in the underdoped regime.

Single crystals of Ba(Fe$_{1-x}$Ru$_x$)$_2$As$_2$ and
BaFe$_2$(As$_{1-x}$P$_x$)$_2$ were grown according to the methods
described in Refs. \cite{ManJinEom,ZRYe}. The Ru and P doping levels
were determined by energy dispersive X-ray spectroscopy. The sample
was cleaved to a rectangular shape with typical dimensions $\sim$
1.50 $\times$ 0.7 mm$^2$ in the $ab$-plane, and 40 to 80 $\mu$m in
$c$-axis. In-plane thermal conductivity was measured in a dilution
refrigerator, using a standard four-wire steady-state method with
two RuO$_2$ chip thermometers, calibrated {\it in situ} against a
reference RuO$_2$ thermometer. Magnetic fields were applied along
the $c$-axis. To ensure a homogeneous field distribution in the
samples, all fields were applied at temperature above $T_c$.

\begin{figure}
\includegraphics[clip,width=7cm]{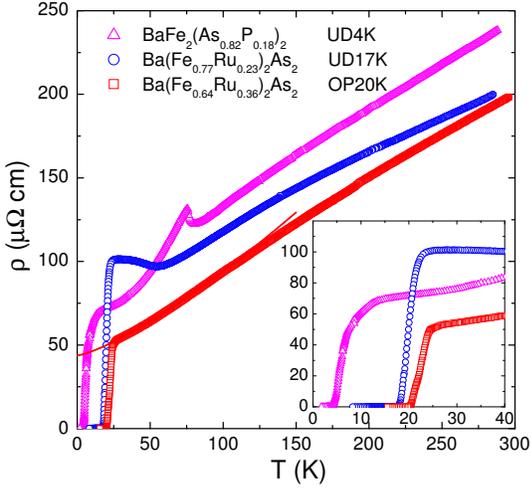}
\caption{(Color online). In-plane resistivity of
Ba(Fe$_{0.64}$Ru$_{0.36}$)$_2$As$_2$,
Ba(Fe$_{0.77}$Ru$_{0.23}$)$_2$As$_2$, and
BaFe$_2$(As$_{0.82}$P$_{0.18}$)$_2$ single crystals. The
low-temperature superconducting transitions are shown in the inset.
Defined by $\rho = 0$, the transition temperatures $T_c$ = 20, 17,
and 4 K are obtained, therefore these three samples are named as
OP20K, UD17K, and UD4K, respectively. The solid line is a fit of the
data between 30 and 90 K to $\rho(T) = \rho_0 +AT^n$ for the OP20K
sample.}
\end{figure}

\begin{figure}
\includegraphics[clip,width=5.5cm]{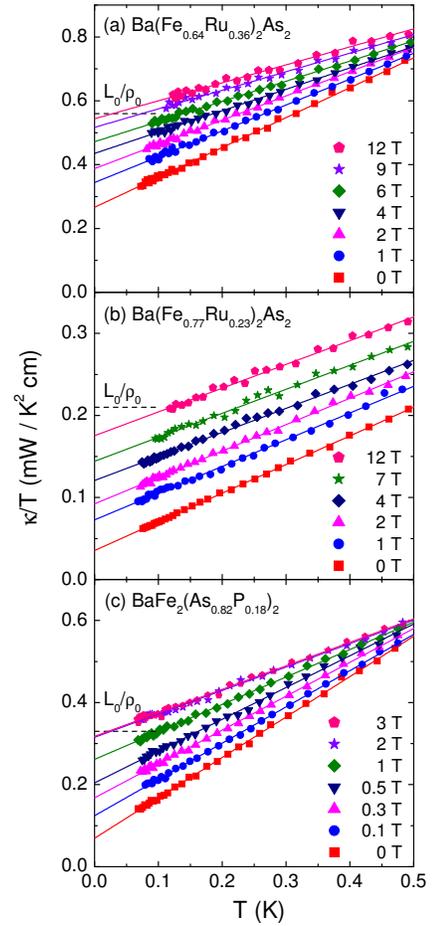}
\caption{(Color online). Low-temperature in-plane thermal
conductivity of Ba(Fe$_{0.64}$Ru$_{0.36}$)$_2$As$_2$,
Ba(Fe$_{0.77}$Ru$_{0.23}$)$_2$As$_2$, and
BaFe$_2$(As$_{0.82}$P$_{0.18}$)$_2$ in zero and magnetic fields
applied along the $c$-axis. The solid lines are $\kappa/T = a + bT$
fit to all the curves, respectively. The dash lines are the
normal-state Wiedemann-Franz law expectation $L_0$/$\rho_0$, with
$L_0$ the Lorenz number 2.45 $\times$ 10$^{-8}$ W$\Omega$K$^{-2}$
and normal-state $\rho_0$ = 44, 115, and 74 $\mu\Omega$cm,
respectively.}
\end{figure}

Fig. 2 shows the in-plane resistivity $\rho(T)$ of
Ba(Fe$_{0.64}$Ru$_{0.36}$)$_2$As$_2$,
Ba(Fe$_{0.77}$Ru$_{0.23}$)$_2$As$_2$, and
BaFe$_2$(As$_{0.82}$P$_{0.18}$)$_2$ single crystals. The transition
temperatures defined by $\rho = 0$ are $T_c$ = 20, 17, and 4 K,
therefore we name these three samples as OP20K, UD17K, and UD4K,
respectively. One can see clear resistivity anomalies in UD4K and
UD17K, but not in OP20K, which manifest the gradual suppression of
spin-density-wave (SDW) order upon P or Ru doping
\cite{SKasahara,FRullier-Albenque,AThaler}. For OP20K, the
resistivity data between 30 and 90 K are fitted to $\rho(T) = \rho_0
+AT^n$, which gives a residual resistivity $\rho_0$ = 43.7 $\pm$ 0.1
$\mu\Omega$cm and $n$ = 1.31 $\pm$ 0.1. Such a non-Fermi-liquid
temperature dependence of $\rho(T)$ is similar to that observed in
BaFe$_2$(As$_{1-x}$P$_x$)$_2$ near optimal doping, which may reflect
the presence of antiferromagnetic spin fluctuations near a quantum
critical point \cite{SKasahara}.

The resistivity of these samples were also measured in magnetic
fields, the highest up to 14.5 T, in order to determine their upper
critical field $H_{c2}$ and normal-state $\rho_0$. For OP20K, UD17K,
and UD4K, we estimate $H_{c2}$ = 23, 19, and 5 T, and normal-state
$\rho_0$ = 44, 115, and 74 $\mu\Omega$cm, respectively.

Fig. 3 shows the temperature dependence of the in-plane thermal
conductivity for OP20K, UD17K, and UD4K in zero and magnetic fields,
plotted as $\kappa/T$ vs $T$. All the curves are roughly linear, as
previously observed in BaFe$_{1.9}$Ni$_{0.1}$As$_2$ \cite{LDing},
KFe$_2$As$_2$ \cite{JKDong1}, and overdoped
Ba(Fe$_{1-x}$Co$_x$)$_2$As$_2$ single crystals
\cite{Tanatar,JKDong2}. Therefore we fit all the curves to
$\kappa/T$ = $a + bT^{\alpha-1}$ with $\alpha$ fixed to 2. The two
terms $aT$ and $bT^{\alpha}$ represent contributions from electrons
and phonons, respectively. Here we only focus on the electronic
term.

For OP20K in zero field, the fitting gives a residual linear term
$\kappa_0/T$ = 0.266 $\pm$ 0.002 mW K$^{-2}$ cm$^{-1}$. This value
is more than 40\% of the normal-state Wiedemann-Franz law
expectation $\kappa_{N0}/T$ = $L_0$/$\rho_0$ = 0.56 mW K$^{-2}$
cm$^{-1}$, with $L_0$ the Lorenz number 2.45 $\times$ 10$^{-8}$
W$\Omega$K$^{-2}$ and normal-state $\rho_0$ = 44 $\mu\Omega$cm. For
optimally doped BaFe$_2$(As$_{0.67}$P$_{0.33}$)$_2$ single crystal,
similar value of $\kappa_0/T \approx$ 0.25 mW K$^{-2}$ cm$^{-1}$ was
obtained, which is about 30\% of its normal-state $\kappa_{N0}/T$
\cite{KHashimoto2}. The significant $\kappa_0/T$ of
Ba(Fe$_{0.64}$Ru$_{0.36}$)$_2$As$_2$ in zero field is attributed to
nodal quasiparticles, which is a strong evidence for nodes in the
superconducting gap \cite{Shakeripour}.

With decreasing doping level, for UD17K and UD4K, $\kappa_0/T$ = 35
$\pm$ 1 and 69 $\pm$ 1  $\mu$W K$^{-2}$ cm$^{-1}$ are obtained, as
seen in Figs. 3(b) and 3(c). These values are about 17\% and 22\% of
their normal-state $\kappa_{N0}/T$, respectively. We note Hashimoto
{\it et al.} already mentioned that the superconducting gap in
overdoped SrFe$_2$(As$_{0.6}$P$_{0.4}$)$_2$ and
BaFe$_2$(As$_{0.36}$P$_{0.64}$)$_2$ has nodes \cite{KHashimoto3}.
Therefore, our observation of significant $\kappa_0/T$ in the
underdoped regime, particularly in the heavily underdoped
BaFe$_2$(As$_{0.82}$P$_{0.18}$)$_2$, further shows the robustness of
nodal superconductivity against doping in P- and Ru-doped 122 iron
pnictides.

The field dependence of $\kappa_0/T$ can provide further support for
the gap nodes \cite{Shakeripour}. In Fig. 4, the normalized
$(\kappa_0/T)/(\kappa_{N0}/T)$ of OP20K, UD17K, and UD4K are plotted
as a function of $H/H_{c2}$. For UD4K, $\kappa/T$ saturates above
$H$ = 3 T, as seen in Fig. 3(c), which is determined as its bulk
$H_{c2}$. For OP20K and UD17K, we use their $H_{c2}$ estimated from
resistivity measurements. To choose a slightly different bulk
$H_{c2}$ does not affect our discussion on the field dependence of
$\kappa_0/T$. Similar data of an overdoped $d$-wave cuprate
superconductor Tl-2201 \cite{Proust}, and
BaFe$_2$(As$_{0.67}$P$_{0.33}$)$_2$ \cite{KHashimoto2} are also
plotted for comparison.

For a nodal superconductor such as Tl-2201 in magnetic field,
delocalized states exist out the vortex cores and dominate the heat
transport in the vortex state, in contrast to the $s$-wave
superconductor. At low field, the Doppler shift due to superfluid
flow around the vortices will yield an $H^{1/2}$ growth in
quasiparticle density of states (the Volovik effect \cite{Volovik}),
thus the $H^{1/2}$ field dependence of $\kappa_0/T$. From Fig. 4,
the behavior of $\kappa_0(H)/T$ in OP20K, UD17K, and UD4K clearly
mimics that in Tl-2201 and BaFe$_2$(As$_{0.67}$P$_{0.33}$)$_2$. In
the inset of Fig. 4, the $\kappa_0(H)/T$ of OP20K, UD17K, and UD4K
obey the $H^{1/2}$ dependence at low field, which supports the
existence of gap nodes.

\begin{figure}
\includegraphics[clip,width=7cm]{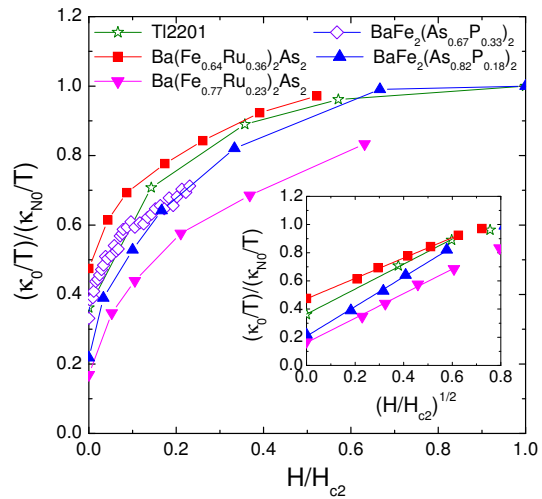}
\caption{(Color online). Normalized residual linear term
$\kappa_0/T$ of Ba(Fe$_{0.64}$Ru$_{0.36}$)$_2$As$_2$,
Ba(Fe$_{0.77}$Ru$_{0.23}$)$_2$As$_2$, and
BaFe$_2$(As$_{0.82}$P$_{0.18}$)$_2$ as a function of $H/H_{c2}$.
Similar data of an overdoped $d$-wave cuprate superconductor Tl-2201
\cite{Proust}, and BaFe$_2$(As$_{0.67}$P$_{0.33}$)$_2$
\cite{KHashimoto2} are also shown for comparison. The behaviors of
$\kappa_0(H)/T$ in OP20K, UD17K, and UD4K clearly mimic that in
Tl-2201 and BaFe$_2$(As$_{0.67}$P$_{0.33}$)$_2$. Inset: the same
data of OP20K, UD17K, UD4K, and Tl-2201 plotted against
$(H/H_{c2})^{1/2}$. The lines represent the $H^{1/2}$ dependence.}
\end{figure}

To our knowledge, so far there are five iron-based superconductors
displaying nodal superconductivity, including KFe$_2$As$_2$
\cite{JKDong1,KHashimoto1}, underdoped Ba$_{1-x}$K$_x$Fe$_2$As$_2$
($x <$ 0.16) \cite{JReid1}, BaFe$_2$(As$_{1-x}$P$_x$)$_2$
\cite{YNakai,KHashimoto2}, LaFePO \cite{Fletcher,Hicks,MYamashida},
and LiFeP \cite{KHashimoto3}. Here we only consider the ``in-plane
nodes", not counting the ``$c$-axis nodes" in underdoped and
overdoped Ba(Fe$_{1-x}$Co$_x$)$_2$As$_2$ as suggested by $c$-axis
heat transport experiments \cite{JReid2}. For the extremely
hole-doped KFe$_2$As$_2$, the nodal gap may have $d$-wave symmetry,
and result from the direct intra-pocket interaction via
antiferromagnetic fluctuations, due to the lack of electron pockets
\cite{JKDong1,RThomale2,SMaiti}. For underdoped
Ba$_{1-x}$K$_x$Fe$_2$As$_2$, it is still not clear how the
superconducting gap transforms from nodeless to nodal at $x \approx
0.16$ \cite{JReid1}. The rest three compounds,
BaFe$_2$(As$_{1-x}$P$_x$)$_2$, LaFePO, and LiFeP, have stimulated
various interpretations of the effect of isovalent P doping
\cite{KKuroki,FaWang2,RThomale,KSuzuki}, and may represent a
peculiar superconducting mechanism.

Our new finding of nodal superconductivity in
Ba(Fe$_{1-x}$Ru$_x$)$_2$As$_2$ reveals the similarity between the
isovalently Ru- and P-doped iron pnictides. In this sense, the P
doping is not that special for inducing nodal superconductivity now,
and the puzzle of P doping in iron-based superconductors has been
partially unwrapped. What next one needs to do is to find out the
common origin for the nodal superconductivity in these isovalently
doped iron pnictides.

Due to the smaller size of P ion than As ion, one common structural
feature of the P-doped compounds is the decrease of pnictogen height
$h_{Pn}$ and increase of As-Fe-As angle
\cite{MTegel,SKasahara,SJiang}. The substitution of larger Ru ion
for Fe ion in Ba(Fe$_{1-x}$Ru$_x$)$_2$As$_2$ results in the increase
of $a$ lattice parameter and decrease of $c$ lattice parameter, thus
the decrease of pnictogen height and increase of As-Fe-As angle
\cite{FRullier-Albenque}. Therefore, both the P and Ru dopants cause
the same trend of structure change in iron arsenides.

With such structure change, the Fermi surface (FS) evolution upon
isovlant P and Ru doping is rather delicate. The main feature, hole
pockets around Brillouin zone (BZ) center and electron pockets
around BZ corners, remains in LaFePO \cite{DHLu},
BaFe$_2$(As$_{1-x}$P$_x$)$_2$ \cite{TShimojima,ZRYe,YZhang},
Ba(Fe$_{1-x}$Ru$_x$)$_2$As$_2$ \cite{VBrouet,RSDhaka}, and LiFeP
\cite{CPutzke}. For LaFePO, Kuroki {\it et al.} have attributed the
low-$T_c$ nodal pairing to the lack of Fermi surface $\gamma$ around
($\pi,\pi$) in the unfolded Brillouin zone, due to the low pnictogen
height \cite{KKuroki}. For BaFe$_2$(As$_{1-x}$P$_x$)$_2$, Suzuki
{\it et al.} have proposed three-dimensional nodal structure in the
largely warped hole Fermi surface and no nodes on the electron Fermi
surface \cite{KSuzuki}. This is supported by recent ARPES
experiments, which found nodal gap in the expanded $\alpha$ hole
pocket at $k_z = \pi$ in BaFe$_2$(As$_{0.7}$P$_{0.3}$)$_2$
\cite{YZhang}, however, it conflicts with earlier ARPES results
which have constrained the nodes on the electron pockets
\cite{TShimojima}. Since ARPES experiments did not find significant
changes in the shape of the Fermi surface or in the Fermi velocity
over a wide range of doping levels in
Ba(Fe$_{1-x}$Ru$_x$)$_2$As$_2$, Dhaka {\it et al.} speculated that
its superconducting mechanism relies on magnetic dilution which
leads to the reduction of the effective Stoner enhancement
\cite{RSDhaka}. In LiFeP, the middle hole pocket has significantly
lower mass enhancement than the other pockets, which implies that
the electron-hole scatter rate is suppress for this pocket and may
result in the lower $T_c$ and nodal gap \cite{CPutzke}.

While the clues for nodal superconductivity are not very clear from
the FS topology except for KFe$_2$As$_2$, Hashimoto {\it et al.}
gathered the available data for the low-energy quasiparticle
excitations in several iron-pnictide superconductors, and suggested
that there is a threshold value of $h_{Pn} \sim$ 1.33 \AA, below
which all the superconductors exhibit nodal superconducting state
\cite{KHashimoto3}. If this is the case, there may exist a
transition from nodal to nodeless state tuned by $h_{Pn}$ in
underdoped Ba(Fe$_{1-x}$Ru$_x$)$_2$As$_2$ and
BaFe$_2$(As$_{1-x}$P$_x$)$_2$. To test this idea, we estimate
$h_{Pn}$ = 1.317, 1.333, 1.340 \AA\ for OP20K, UD17K, UD4K from the
roughly linear increase of $h_{Pn}$ with decreasing Ru or P doping
\cite{FRullier-Albenque,SKasahara}.

One can see that both $h_{Pn}$ of UD17K and UD4K are slightly larger
than the proposed threshold value 1.33 \AA. In particular, the
$h_{Pn}$ of UD4K is comparable to that of overdoped
Ba(Fe$_{0.89}$Co$_{0.11}$)$_2$As$_2$, which is a nodeless
superconductor \cite{KHashimoto3}. Since our thermal conductivity
data suggest UD17K and UD4K are nodal superconductors, $h_{Pn}$
should not be considered as the only parameter for tuning between
nodeless and nodal superconducting states. By saying this, we do not
deny its importance, since $h_{Pn}$ of the underdoped
Ba(Fe$_{1-x}$Ru$_x$)$_2$As$_2$ and BaFe$_2$(As$_{1-x}$P$_x$)$_2$ are
still very close to the threshold value 1.33 \AA. More careful
considerations of the structural parameters, FS topology, and local
interactions are needed to clarify the origin of the nodal
superconductivity in isovalently doped iron pnictides.

In summary, we have measured the thermal conductivity of
Ba(Fe$_{0.64}$Ru$_{0.36}$)$_2$As$_2$,
Ba(Fe$_{0.77}$Ru$_{0.23}$)$_2$As$_2$, and
BaFe$_2$(As$_{0.82}$P$_{0.18}$)$_2$ single crystals down to 50 mK. A
significant $\kappa_0/T$ at zero field and an $H^{1/2}$ field
dependence of $\kappa_0(H)/T$ at low field give strong evidences for
nodal superconductivity in all three compounds. Comparing with
previous P-doped iron pnictides, our new finding suggest that the
nodal superconductivity induced by isovalent Ru and P doping may
have the same origin. With decreasing doping level, nodal
superconducting state persists robustly in heavily underdoped
BaFe$_2$(As$_{0.82}$P$_{0.18}$)$_2$, suggesting that the $h_{Pn}$ is
not the only tuning parameter, thus putting constraint on
theoretical models. Finding out the origin of these nodal
superconducting states will be crucial for getting a complete
electronic pairing mechanism in the iron-based high-$T_c$
superconductors.

This work is supported by the Natural Science Foundation of China,
the Ministry of Science and Technology of China (National Basic
Research Program No: 2009CB929203 and 2012CB821402), Program for
Professor of Special Appointment (Eastern Scholar) at Shanghai
Institutions of Higher Learning, and STCSM of China. The work at
Postech was supported by Leading Foreign Research Institute
Recruitment Program(2010-00471), Basic Science Research Programs
(2010-0005669) through the National Research Foundation of Korea(NRF) \\

$^*$ E-mail: shiyan$\_$li@fudan.edu.cn

\end{document}